\documentclass[conference]{IEEEtran}
\IEEEoverridecommandlockouts
\usepackage{amsmath}
\usepackage{amssymb}
\usepackage[font=footnotesize]{caption}
\usepackage{mathrsfs}
\usepackage{amsfonts}
\usepackage{amssymb}
\usepackage{euscript}
\usepackage{graphicx}
\DeclareMathOperator*{\argmax}{arg\,max}
\DeclareMathOperator*{\argmin}{arg\,min}
\usepackage{cite}
\usepackage{tabularx}
\usepackage{csquotes}
\begin{document}
\title{Swish-Driven GoogleNet for Intelligent Analog Beam Selection in Terahertz Beamspace MIMO}
\author{\IEEEauthorblockN {Hosein~Zarini$^{\dag}$, Mohammad Robat Mili$^{\S}$, Mehdi~Rasti$^{\dag, \star\star}$, Sergey Andreev$^{\star}$, and Pedro H. J. Nardelli$^{\star\star}$
}
		$^{\dag}$Department of Computer Engineering, Amirkabir University of Technology, Tehran, Iran\\
		$^\S$Electronics Research Institute, Sharif University of Technology, Tehran, Iran\\
		$^\star$Tampere University, Tampere, Finland\\
		$^{\star\star}$Lappeenranta-Lahti University of Technology, Lappeenranta, Finland
	}
\maketitle

\begin{abstract}
In this paper, we propose an intelligent analog beam selection strategy in a terahertz (THz) band beamspace multiple-input multiple-output (MIMO) system. First inspired by transfer learning, we fine-tune the pre-trained off-the-shelf GoogleNet classifier, to learn analog beam selection as a multi-class mapping problem. Simulation results show 83\% accuracy for the analog beam selection, which subsequently results in 12\% spectral efficiency (SE) gain, upon the existing counterparts.
Towards a more accurate classifier, we replace the conventional rectified linear unit (ReLU) activation function of the GoogleNet with the recently proposed Swish and retrain the fine-tuned GoogleNet to learn analog beam selection. It is numerically indicated that the fine-tuned Swish-driven GoogleNet achieves 86\% accuracy, as well as 18\% improvement in achievable SE, upon the similar schemes.
Eventually, a strong ensembled classifier is developed to learn analog beam selection by sequentially training multiple fine-tuned Swish-driven GoogleNet classifiers. According to the simulations, the strong ensembled model is 90\% accurate and yields 27\% gain in achievable SE, in comparison with prior methods.
\end{abstract}


\section{Introduction}
\par Over the recent years, beamspace technology\cite{beamspace1} has attracted a major attention in high-frequency bands, as an alternative to the conventional massive multiple-input-multiple-output (MIMO) architecture. In latter case, each antenna element requires a specific radio frequency (RF) chain\footnote{RF chains are known as dominant modules in energy consumption, hardware cost and complexity order of conventional massive MIMO systems.}, which makes this architecture inefficient in practice, owing to a massive number of required RF chains.
In beamspace technology nevertheless, the scattered signals of divergent paths (beams) can be concentrated upon a limited number of dominant beams and the spatial domain channel is thereby transformed into the beamspace domain channel. To this reason, from a massive number of beams, merely a limited number is adopted, which in turn necessitates few RF chains for a reliable beam steering. 
\par The hybrid analog-digital beamspace MIMO is consequently a reasonable system in terms of energy, cost, and complexity, provided that the analog beam selection is efficiently performed. Unfortunately, this sets out new challenges due to the massive number of beams. While on one hand, the prior optimization-based analog beam selection efforts such as those in\cite{access} impose expensive computational burden to the transceivers,  the low-complexity machine/deep learning approaches like \cite{access2} and \cite{DT} on the other hand, suffer from accuracy loss in this regard. According to the statistics in \cite{conf}, trained on environmental samples (e.g., the line-of-sight (LoS) and non-line-of-sight (NLoS) beams), two well-known classifiers i.e., the linear SVM \cite{access2} and the decision tree \cite{DT} are only 33\% and 55\% accurate, respectively, which in turn brings about a non-trivial performance loss for the beamspace architecture.
\par The main contribution of this paper is to relieve the precision fall in prior learning-aided works on analog beam selection, by proposing a fine-tuned deep learning technique, along with an ensemble learning technique as follows.
\begin{itemize}
    \item First we account for the analog beam selection problem as a multi-class classification task. To this aim, we retrain the pre-trained off-the-shelf GoogleNet classifier\cite{GoogleNet} based on the concept of transfer learning\cite{transfer1}, so as to learn the analog beam selection. Simulation results verify that the retrained GoogleNet exhibits some 83\% accuracy for the analog beam selection and achieves by up to 12\% gain in achievable spectral efficiency (SE) upon the counterparts, when signal-to-noise-ratio (SNR) is 30dB.
    \item We fine-tune the GoogleNet classifier for a beyond classification precision, by replacing its conventional activation function i.e., the rectified linear unit (ReLU) with the Swish activation function\cite{Swish}. It is numerically shown that retraining the fine-tuned GoogleNet achieves some 86\% accuracy, as well as 18\% achievable SE gain upon the counterparts, at SNR~=~30dB. 
    \item In addition, the performance of the proposed analog beam selection scheme is further enhanced  by sequentially incorporating multitude of the fine-tuned GoogleNets (each one is known as a weak learner) into an ensembled model (known as a strong learner)\cite{gradBoost}. The proposed strong learner according to the simulations outperforms the achievable SE of the prior counterparts, by up to 27\%, while yielding 90\% accuracy, when SNR~=~30dB.
    \end{itemize}
\par In remaining of the paper, Sections II and III describe the system setup and the solution approach, whereas the simulation results and conclusions are
presented in Sections IV and V, respectively.
\begin{figure}
\centering
\includegraphics[width=9.0cm,height=5.0cm]{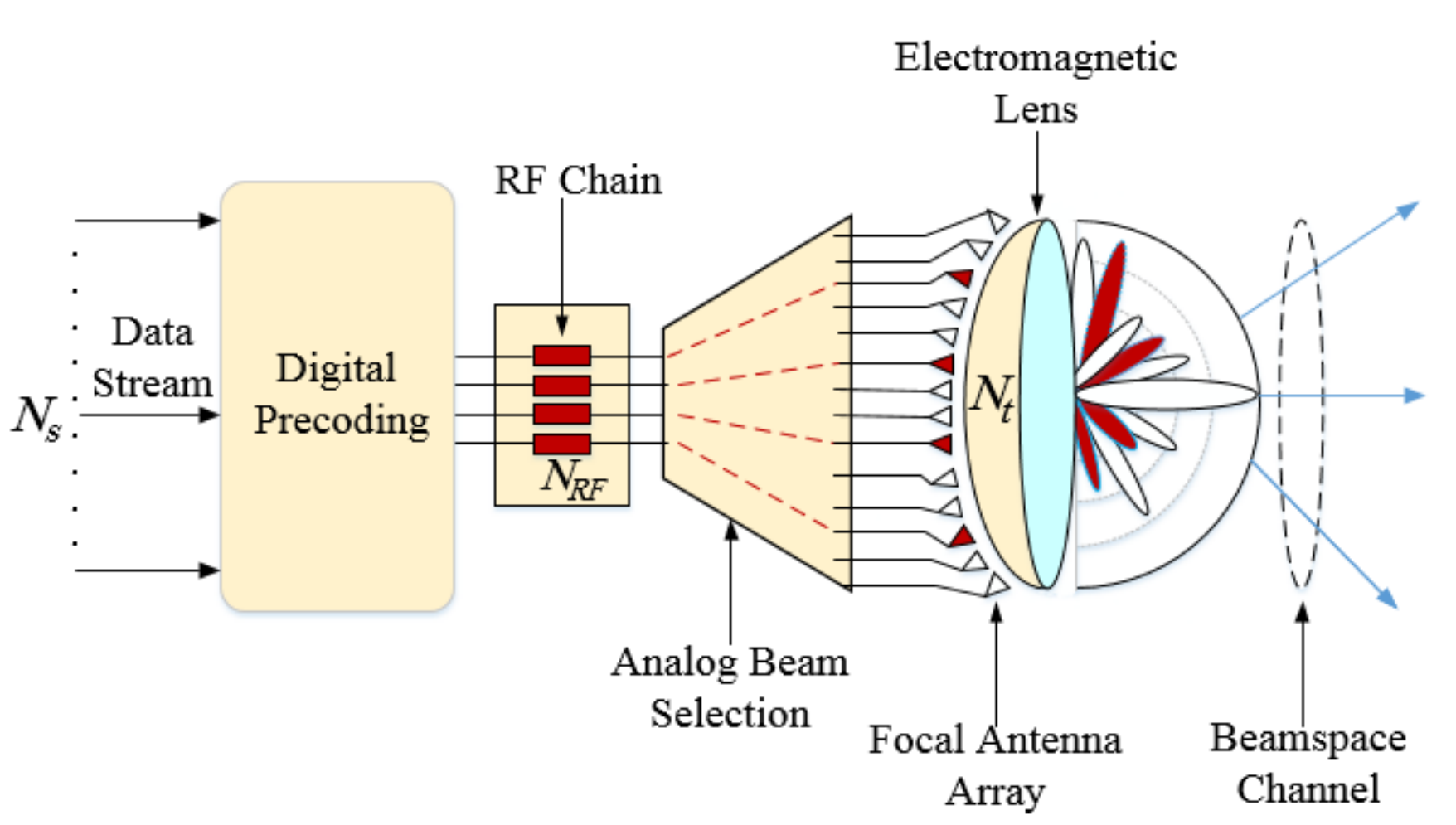}\caption{The hybrid analog-digital beaspace MIMO architecture at the transmitter.}
\label{fig:lens}
\end{figure}
\section{System Setup}
\subsection{Hybrid Analog-Digital Architecture}
Consider a downlink THz communication, where the transmitter employs $N_{\textrm{\textrm{t}}}$($N_{\textrm{\textrm{t}}}^{RF}$) transmit antennas (transmit RF chains) for serving a receiver, equipped with $N_{\textrm{\textrm{r}}}$($N_{\textrm{\textrm{r}}}^{RF}$) receive antennas (receive RF chains). The system multiplexing gain or equivalently, the number of simultaneously communicated data streams is $N_{s}$~=~min$(N_{\textrm{\textrm{t}}}^{RF},N_{\textrm{\textrm{r}}}^{RF})$ and the power-normalized transmit symbols, are denoted by $\mathbf{s}\in
\mathbb{C}^{N_{s}\times1}$, where $\mathbb{E}\left[  \mathbf{s}\mathbf{s}^{H}\right]  =\mathbf{I}_{N_{s}}$. The transceivers enjoy a hybrid analog-digital beamspace architecture to preserve the system flexibility, as well as the efficiency in hardware cost and energy consumption\cite{beamspace1}. As demonstrated, in Fig.~\ref{fig:lens}, a baseband digital matrix $\mathbf{F}_{\textrm{BB}}\in
\mathbb{C}^{N_{\textrm{\textrm{t}}}^{RF}\times N_{s}}$ is leveraged at the transmitter, followed by an analog beam selection network, denoted by $\mathbf{S}_{\textrm{\textrm{t}}}\in{\mathbb{{R}}^{N_{\textrm{\textrm{t}}}\times{N_{\textrm{\textrm{t}}}^{RF}%
}}}$ in matrix form for mapping $N_{\textrm{\textrm{t}}}^{RF}$ transmit RF chains into a subset of $N_{\textrm{\textrm{t}}}$ transmit antennas/beams. Eventually, a lens antenna array is deployed at the transmitter, including an energy-focusing electromagnetic lens, where its focal surface is equipped with a large-scale antenna array. 
\par At the receiver side reversely, once the lens antenna array receives the signals, a mapping is performed between the predominant receive antennas/beams and the receive RF chains through the receive analog beam selection network $\mathbf{S}_{\textrm{\textrm{r}}}$ $\in \mathbb{{R}}^{N_{\textrm{\textrm{r}}}\times N_{\textrm{\textrm{r}}}^{RF}}$, where a baseband digital combining matrix $\mathbf{W}_{\textrm{BB}}$ $\in \mathbb{C}^{N_{\textrm{\textrm{r}}}^{RF}\times N_{s}}$ is embedded afterwards to obtain the transmit symbols. Thus, the discrete-time received baseband complex signal is given by
$
y=\mathbf{W}_{\textrm{BB}}^{H}\mathbf{S}_{\textrm{\textrm{r}}}^{H}\mathbf{H}_{b}\mathbf{x}+\mathbf{W}%
_{\textrm{BB}}^{H}\mathbf{S}_{\textrm{\textrm{r}}}^{H}\mathbf{n}, \label{M2}%
$
wherein $\mathbf{n}$ $\sim N(0,\sigma^{2}\mathbf{I}_{N_{\textrm{\textrm{r}}}})$ is the
additive white Gaussian noise (AWGN) with a noise power $\sigma^{2}$ and $\mathbf{H}_{b}$ denotes the THz beamspace channel. 

\subsection{Communicating THz Channel} 
According to the well-known Saleh-Valenzuela geometric model \cite{Saleh}, a ray-based clustered THz channel is assumed with $N_{{cl}}$ cluster of scatterers, each contributes $N_{{ray}}$ propagation rays. Also, a limited angle-of-departure/arrival (AoD/AoA) spread is supposed for a typical cluster $l$, denoted
by $\psi_{\textrm{t}}^{l}$ and $\psi_{\textrm{r}}^{l}$, respectively. For a typical cluster/ray $l/u$, the complex-valued gain is denoted by $\alpha^{{l,u}}$, while the physical AoD and AoA for the transmitter and receiver is respectively denoted by $\theta_{{\textrm{t}}}^{{l,u}}\in \psi_{{\textrm{t}}}^{{l}}$, and $\theta_{{\textrm{r}}}^{{l,u}}\in \psi_{{\textrm{r}}}^{{l}}$, respectively.
Let us denote the antenna element spacing by $d$, the speed of light by ${c}$, the wavelength by $\lambda=c/f_{c}$, and the carrier frequency by $f_{c}$. Then, the spatial AoD/AoA can be represented by $
\phi_{{\textrm{t}}}^{{l,u}}=(d/\lambda)\sin \theta _{{\textrm{t}}}^{{l,u}}$ and $\phi
_{{\textrm{r}}}^{{l,u}}=(d/\lambda)\sin \theta _{{\textrm{r}}}^{{l,u}},$ respectively. Accordingly, the narrowband discrete-time spatial domain THz channel
$\mathbf{H}%
\in \mathbb{C}^{N_{{\textrm{r}}}\times N_{{\textrm{t}}}}$ is expressed as $\mathbf{H}=\gamma\sum_{{l}=1}^{N_{{cl}}} \sum_{{u}=1}^{N_{{ray}}} \alpha _{{l,u}}\mathbf{a}_{{\textrm{r}}}\left( \phi _{{\textrm{r}}}^{{l,u}}\right)\mathbf{a}%
_{{\textrm{t}}}^{\textrm{H}}\left( \phi _{{\textrm{t}}}^{{l,u}}\right),$
with the normalization factor $\gamma=\sqrt{N_{{\textrm{r}}}N_{{\textrm{t}}}/{N_{{cl}}N_{{ray}}}}.$
Following the uniform linear array (ULA), the antenna array responses at the transmitter/receiver, are represented by $\mathbf{a}_{{\textrm{t}}}\left( \phi _{{\textrm{t}}}^{{l,u}}\right)  =\frac{1}{\sqrt{N_{{\textrm{t}}}}
}\left[ 1,e^{j2\pi \phi _{{\textrm{t}}}^{{l,u}}},...,e^{j2\pi \left( N_{{\textrm{t}}}-1\right)
\phi _{{\textrm{t}}}^{{l,u}}}\right] ^{H} \in \mathbb{C}^{N_{{\textrm{t}}}\times 1}$ and $\mathbf{a}_{{\textrm{r}}}\left( \phi _{{\textrm{r}}}^{{l,u}}\right)  =\frac{1}{\sqrt{N_{{\textrm{r}}}}}
\left[ 1,e^{j2\pi \phi _{{\textrm{r}}}^{{l,u}}},...,e^{j2\pi \left( N_{{\textrm{r}}}-1\right)
\phi _{{\textrm{r}}}^{{l,u}}}\right] ^{H} \in \mathbb{C}^{N_{{\textrm{r}}}\times 1}$, respectively. Important to note that the THz channel $\textbf{H}$ in spatial domain is effectively transformed into the equivalent channel in beamspace domain $\textbf{H}_b$, on the basis of DFT operations in lens antenna array (see\cite{wideband} for details).
\subsection{Problem Statement}
\par In the considered hybrid analog-digital beamspace massive MIMO system, we focus on achieving analog beam selection for the transmitter and receiver $\mathbf{S}_{{\textrm{t}}},\mathbf{S}_{{\textrm{r}}}$, under the assumption of given the precoding/combining matrices and given the beamspace channel. This problem can be formally stated as\cite{surv}
\begin{align}\label{opt1}
    &\min_{\mathbf{S}_{{\textrm{t}}},\mathbf{S}_{{\textrm{r}}}}~||\mathbf{H}_{{b}}-\mathbf{S}_{{\textrm{r}}}\mathbf{W}_{\textrm{BB}}\mathbf{F}_{\textrm{BB}}^{H}\mathbf{S}_{{\textrm{t}}}^{H}||^{2}
    \\&\nonumber s.t.
    \\&\nonumber\mathbf{S}_{{\textrm{r}}}\in{\mathbf{\mathcal{S}}_{{\textrm{r}}}},
    \\&\nonumber \mathbf{S}_{{\textrm{t}}}\in{\mathbf{\mathcal{S}}_{{\textrm{t}}}},
\end{align}
where ${\mathbf{\mathcal{S}}_{{\textrm{t}}}}$ and ${\mathbf{\mathcal{S}}_{{\textrm{r}}}}$ are the analog beam selection candidate sets at the transmitter and receiver, respectively. The optimal solution for acquiring the analog beam selection variables $\mathbf{S}_{{\textrm{r}}}$ and $\mathbf{S}_{{\textrm{t}}}$ is the exhaustive search method, which is computationally expensive and definitely infeasible for a beamspace massive MIMO system. 
\begin{figure*}
	\centering
		\includegraphics[width=13cm,height=6.0cm]{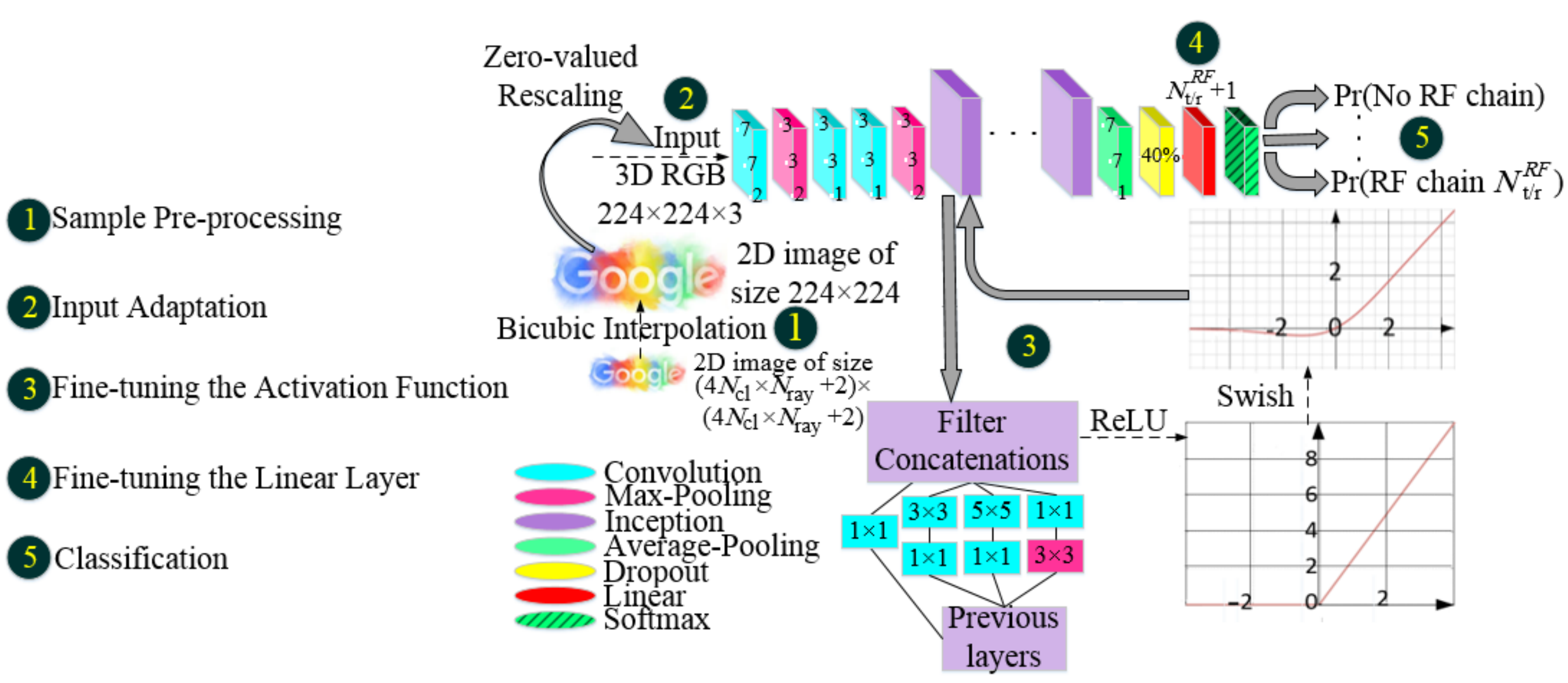}
		\caption{The architecture of GoogleNet, modifications performed on training samples to fit into the input layer, replacing ReLU with Swish and setting the number of linear layer classes from 1000 into $N^{RF}_{\textrm{t}}$+1 (or $N^{RF}_{\textrm{r}}$+1).}
		\label{fig:GoogleNet}
\end{figure*}
\section{Solution Approach}
In this section, the training sample set acquisition, the Swish-driven GoogleNet, transfer learning and ensemble learning are respectively elaborated as our solution approach to (\ref{opt1}).
\subsection{Sample Set Acquisition}
We consider the network parameters path gain, transmit power, AoA and AoD constituting $4N_{{cl}}\times N_{{ray}}+2$ random
real-valued features with one feature for the transmit power of the transmitter, one feature for the path gain, $2N_{{cl}}\times N_{{ray}}$ features for the AoDs/AoAs of the transmitter/receiver, and as such $2N_{{cl}}\times N_{{ray}}$ features for the real and imaginary parts of the complex-valued gain to form a data sample. In following, we conduct a normalization process, a Gaussian mixture model (GMM) fitting, and a labeling operation over the samples. 
\subsubsection{Normalization}
Due to the diversity in sample ranges (e.g., the transmit power is based on dB, while AoDs are within [0,2$\pi$]), a normalization pre-processing needs to be accomplished for each feature of samples as $\bar{a}_{f}^{m}=\big[a_{f}^{m}-\mathrm{Mean}(a_{f}^{m})\big]\times\big[{a_{f}^\textrm{max}-a_{f}^\textrm{min}}\big]^{-1},$
where $a_{f}^{m}$ indicates the value of $f$th feature in $m$th sample and $\textrm{Mean}(a_{f}^{m})$ is the mean of all $a_{f}^{m}$. Besides, $a_{f}^\textrm{max}$ and $a_{f}^\textrm{min}$ denote the maximum and minimum values of the $f$th feature among all samples, respectively. Hence, the $m$th sample as a feature row vector can be characterized as $\textrm{z}_m\in{\mathbb{C}^{1\times (4N_{{cl}}\times N_{{ray}}+2}})$ with $4N_{{cl}}\times N_{{ray}}+2$ normalized features. 
\subsubsection{GMM Fitting}
\par Since the beamspace channel features $\phi_{t}$, $\phi_{r}$ and $\alpha$ follow a Gaussian distribution\cite{LAMP2}, we adopt a GMM for appropriately fitting the beamspace channel. In doing so, we have
$\mathbf{\tilde{H}}_{b} =
A\times
    \Big(\sum_{k=1}^{K}w_{k}\textrm{exp}\Big(-\frac{(\phi _{\textrm{r}}-\mu_{{\phi _{\textrm{r}}}_{k}})^{2}}{2\sigma^{2}_{{\phi _{\textrm{r}}}_{k}}}-\frac{(\phi _{\textrm{t}}-\mu_{{\phi _{\textrm{t}}}_{k}})^{2}}{2\sigma^{2}_{{\phi _{\textrm{t}}}_{k}}}-\frac{(\phi _{\textrm{r}}-\mu_{{\alpha}_{k}})^{2}}{2\sigma^{2}_{{\alpha}_{k}}}
    \Big)\Big),$ with the GMM-fitted beamspace channel $\mathbf{\tilde{H}}_{b}$, the GMM amplitude $A$, and $K$ Gaussian components, where $w_{k}\in{[0,1]}$ is the weight of the Gaussian component $k$ and $\sum_{k=1}^{K}w_{k}=1$. Note that in $\mathbf{\tilde{H}}_{b}$, the central coordinates are ($\mu_{{\phi _{\textrm{r}}}_{k}},\mu_{{\phi _{\textrm{t}}}_{k}},\mu_{{\alpha}_{k}}$), whereas $\sigma_{{\phi _{\textrm{r}}}_{k}}$, $\sigma_{{\phi _{\textrm{t}}}_{k}}$ and $\sigma_{{\alpha}_{k}}$ indicate their corresponding standard deviation. In vector representation, the Gaussian component $k$ can be expressed as $q_{k}=[w_{k},\mu_{{\phi _{\textrm{r}}}_{k}},\mu_{{\phi _{\textrm{t}}}_{k}},\mu_{{\alpha}_{k}},\sigma_{{\phi _{\textrm{r}}}_{k}},\sigma_{{\phi _{\textrm{t}}}_{k}},\sigma_{{\alpha}_{k}}]$.
Equivalently, the spatial features of the samples based on all of the Gaussian components can be given by $\textbf{\textrm{q}}=[A;q_{1};q_{2};...;q_{K}]^{T}=[A,\mu_{{\phi _{\textrm{r}}}_{1}},\mu_{{\phi _{\textrm{t}}}_{1}},\mu_{{\alpha}_{1}},\sigma_{{\phi _{\textrm{r}}}_{1}},\sigma_{{\phi _{\textrm{t}}}_{1}},\sigma_{{\alpha}_{1}},\mu_{{\phi _{\textrm{r}}}_{2}},\mu_{{\phi _{\textrm{t}}}_{2}},\mu_{{\alpha}_{2}},\sigma_{{\phi _{\textrm{r}}}_{2}},\sigma_{{\phi _{\textrm{t}}}_{2}},$
$\sigma_{{\alpha}_{2}},...,\mu_{{\phi _{\textrm{r}}}_{K}},\mu_{{\phi _{\textrm{t}}}_{K}},\mu_{{\alpha}_{K}},\sigma_{{\phi _{\textrm{r}}}_{K}},\sigma_{{\phi _{\textrm{t}}}_{K}},\sigma_{{\alpha}_{K}}]^{T}$. Finally, the optimal vector $\textbf{\textrm{q}}$, which is used to model the beamspace channel distribution can be determined according to \cite{GMM}.
\subsubsection{Labeling}
\par 
The cost function for evaluating the analog beam selection decisions (i.e., labeling) is the objective in (\ref{opt1}), which equivalently optimizes the achievable SE\cite{surv}. The labeling phase is a multi-class mapping operation that determines the optimum (beam,RF) candidates obtained from\cite{sample}, wherein each RF chain is a class label to which, analog beams are assigned to.
\subsection{GoogleNet Architecture}
As an off-the-shelf pre-trained network, GoogleNet has been trained by the well-known datasets (e.g., ImageNet) beforehand, while its weights, biases, and other training parameters have been already set. 
According to Fig.~\ref{fig:GoogleNet}, the network has 22 layers with an input layer of size 224$\times$224$\times$3 for receiving a two-dimensional (2D) image of width and length 224 and 3 channels of RGB (i.e., red, green, and blue). The main parts in GoogleNet architecture are its inception modules that incorporate multiple convolutions, kernels, and max-pooling layers, simultaneously within a single layer. 
The main activation function in GoogleNet is ReLU, which is computationally cheap and embedded upon a filter concatenation layer within the inception module (see Fig.~\ref{fig:GoogleNet}) for improved training performance. By going deeper in GoogleNet architecture as observed in Fig.~\ref{fig:GoogleNet}, the linear layer of size 1000 is followed by a dropout layer with $40\%$ ratio of dropped outputs and connected to a Softmax activation function with 1000 classes.

\subsection{Swish-driven GoogleNet}
\par Despite its accurate classification capability, the performance of GoogleNet can still be improved by minor architectural modifications. For instance, the authors in \cite{modification1} proposed to substitute the ReLU activation functions in GoogleNet with the Leaky-ReLU (an extension of the conventional ReLU) for faster convergence. In \cite{modification2}, the large convolutional filters in GoogleNet were factorized into smaller ones, and this modification benefited for the middle layers of GoogleNet. In this paper, we modify the ReLU activation functions in filter concatenation layer of the inception modules (see Fig.~\ref{fig:GoogleNet}) in GoogleNet architecture by the Swish\cite{Swish}. The latter is a self-gated, smooth, and non-monotonic activation function recently proposed by Google Brain Team. By definition, the Swish activation function for an any input $x$ can be given by $f^\textrm{Swish}(x) = x.f^\textrm{Sigmoid}(x) = \frac{x}{1+e^{-x}}.$
The numerical results in \cite{Swish} indicate that the Swish is more precise than the ReLU (and its alternative extensions, such as Leaky-ReLU) with a similar level of computational complexity, especially in very deep architectures.
\subsection{Transfer Learning}
\par To fit the size of samples into the input layer of the fine-tuned Swish-driven GoogleNet, certain modifications need to be necessarily accomplished in accordance with Fig.~\ref{fig:GoogleNet}.
First, we extend the dimensionality of a typical sample $\textrm{z}_m$ of size $(4N_{{cl}}\times N_{{ray}}+2)$ into a matrix form of size $(4N_{{cl}}\times N_{{ray}}+2)\times(4N_{{cl}}\times N_{{ray}}+2)$ as a 2D image. Next, we preform an image resizing through the interpolation technique to transform each sample into the size of $224\times224$. Specifically, we use bicubic interpolation that can preserve the quality of the primary image by extracting the most determinant properties (which correspondingly are related to the most dominant features of the sample in our case).
The $224\times224$ resized 2D image of $\textrm{z}_m$ is eventually extended into a three dimensional (3D) image by using zero-valued rescaling. To do so, the RGB color triplet for each pixel is set to zero, thus leading to a 3D RGB image of size $224\times224\times3$ to feed the input layer of the GoogleNet.
\par We further fine-tune the final linear layer of the GoogleNet by setting $N_{\textrm{t}}^{RF}$+1 classes for the transmitter (or $N_{\textrm{r}}^{RF}$+1 for the receiver), which trains the GoogleNet to map any sample (beam) into the correct class (RF chain). During the training process, the beamspace channel feature space is processed through the layers of the GoogleNet, while its main features (energy-focused features of the beam) are extracted. The Softmax classifier eventually learns a multi-class mapping based on the labeled samples obtained from\cite{sample}. 
The probability of the $i$th RF chain being selected by the Softmax function is $\delta(N_{\textrm{t}}^{RF})_{i}=\big[e^{\big({N_{\textrm{t}}^{RF}}\big)_i}\big]\times\big[\sum_{i=1}^{|N_{\textrm{t}}^{RF}|}e^{\big({N_{\textrm{t}}^{RF}}\big)_i}\big]^{-1}.$
\par Finally, as observed in Fig.~\ref{fig:GoogleNet}, a modified version of the GoogleNet is trained by fine-tuning its linear layer and activation functions. This approach is known as transfer learning, whereby the main layers of a pre-trained network are directly imported into the new application, while other layers remain unchanged. By doing so, the fine-tuned GoogleNet learns analog beam selection at the transceivers based on the beamspace channel feature space, while its internal weights, biases, and other parameters are mainly fixed. 
\begin{figure}
\centering\includegraphics[width=9.0cm,height=5.50cm]{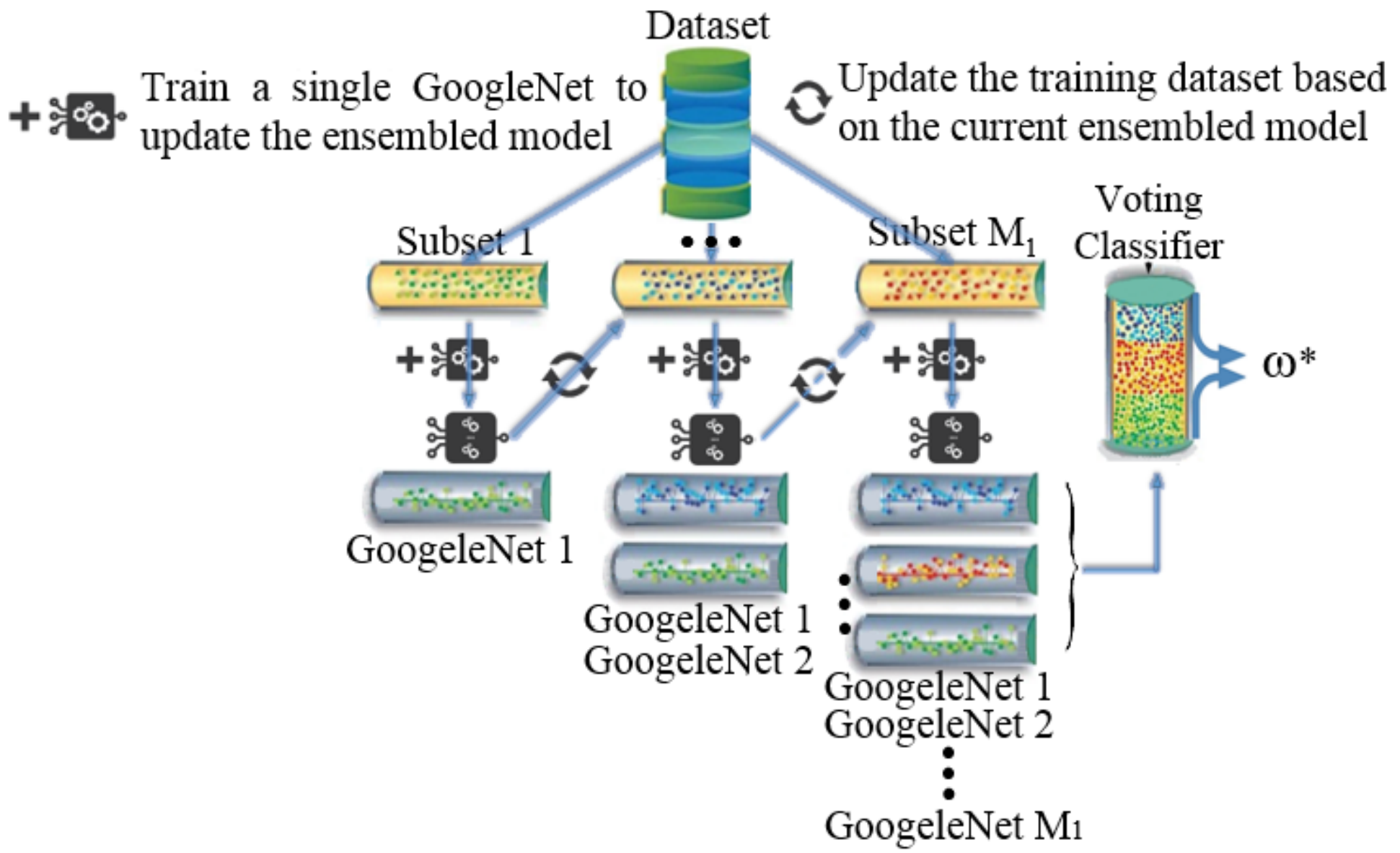}
\caption{Ensemble learning schematic.}
\label{fig:ensemble}
\end{figure}
\begin{figure*}
	\centering
	\begin{tabular}{lccccc}
	\includegraphics[width=6.0cm,height=4.0cm]{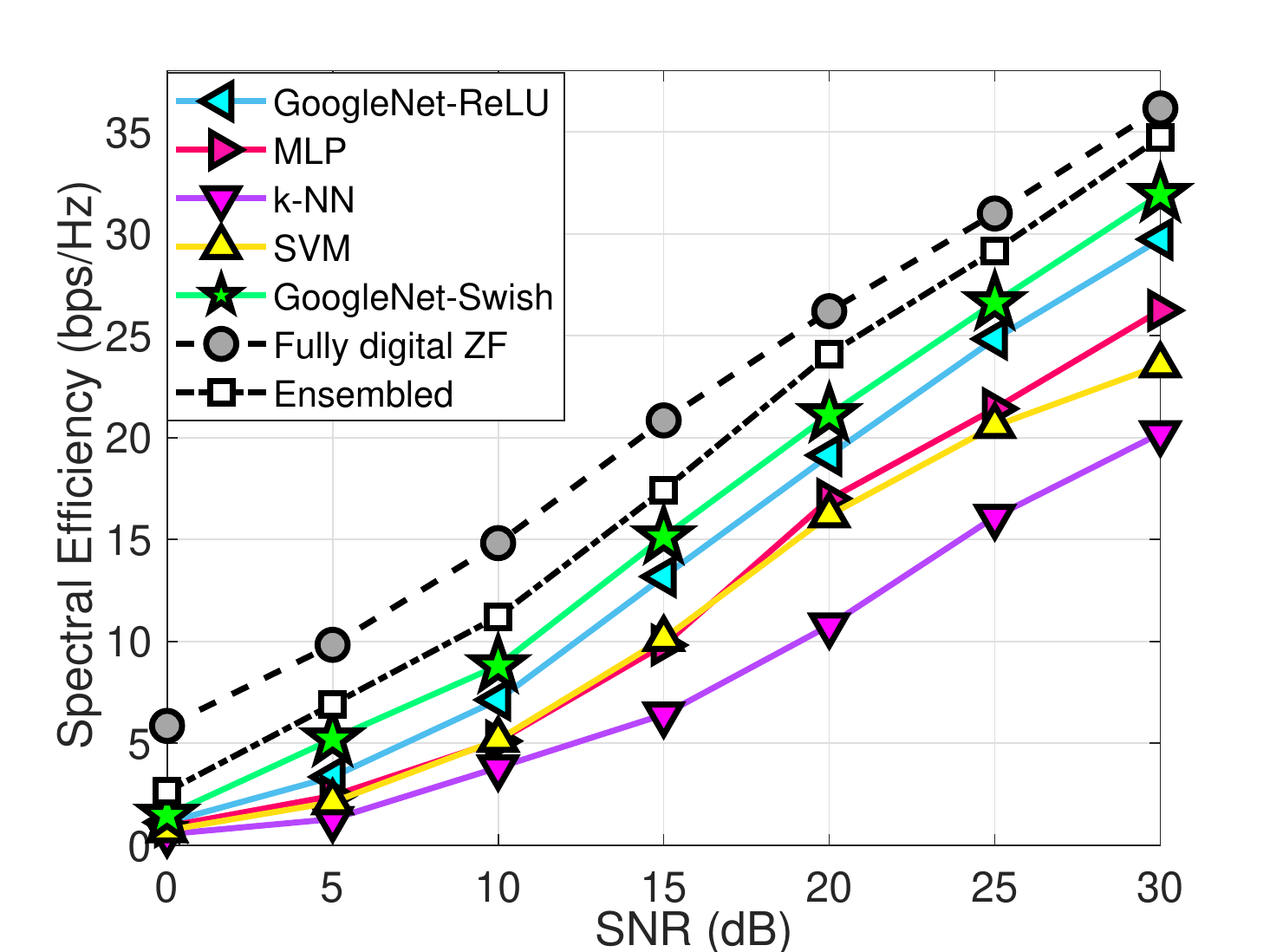}&\includegraphics[width=11.0cm,height=3.90cm]{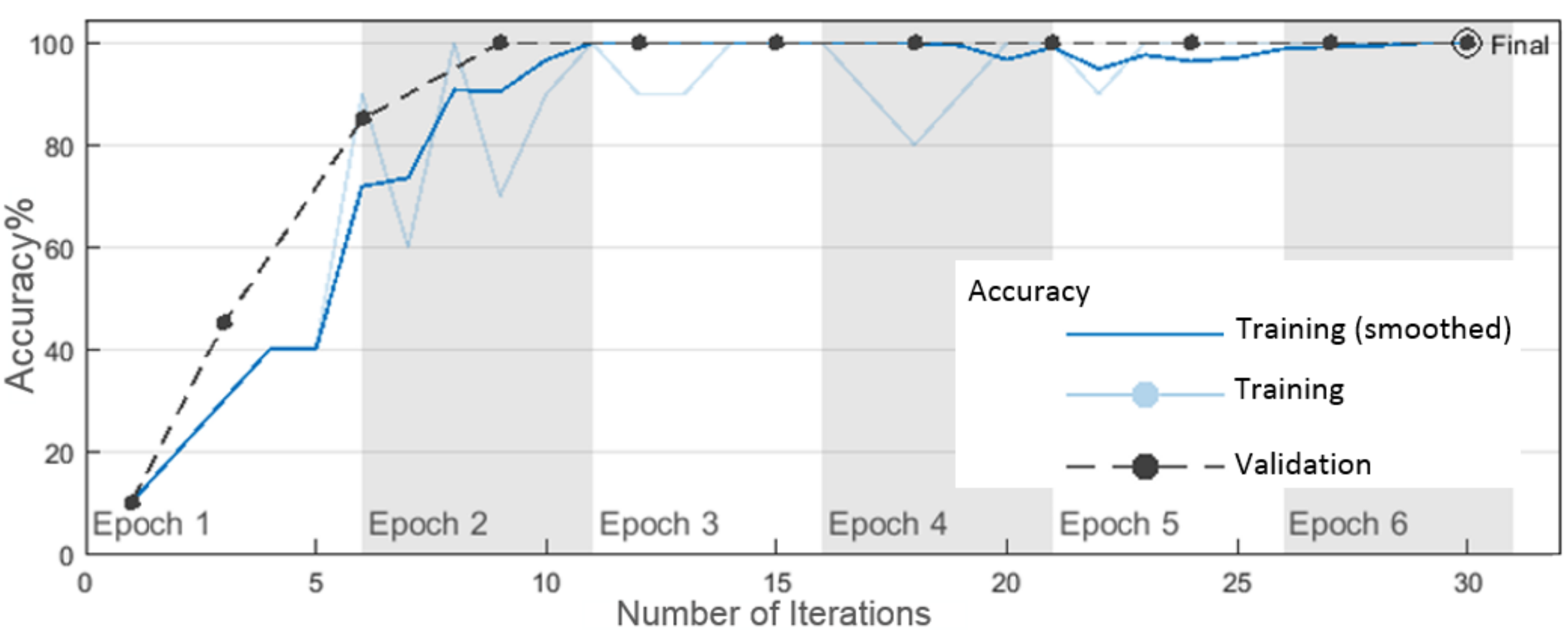}\\
	\hspace{0.50cm}(a) Achievable SE vs. varying SNR & \hspace{0.3cm} (b) The convergence of the Swish-driven GoogleNet (accuracy)
	\end{tabular}
\end{figure*}
\begin{figure*}
\hspace{0.0cm}
	\centering
	\begin{tabular}{lccccc}
	\hspace{-0.3cm}\includegraphics[width=6.0cm,height=4.0cm]{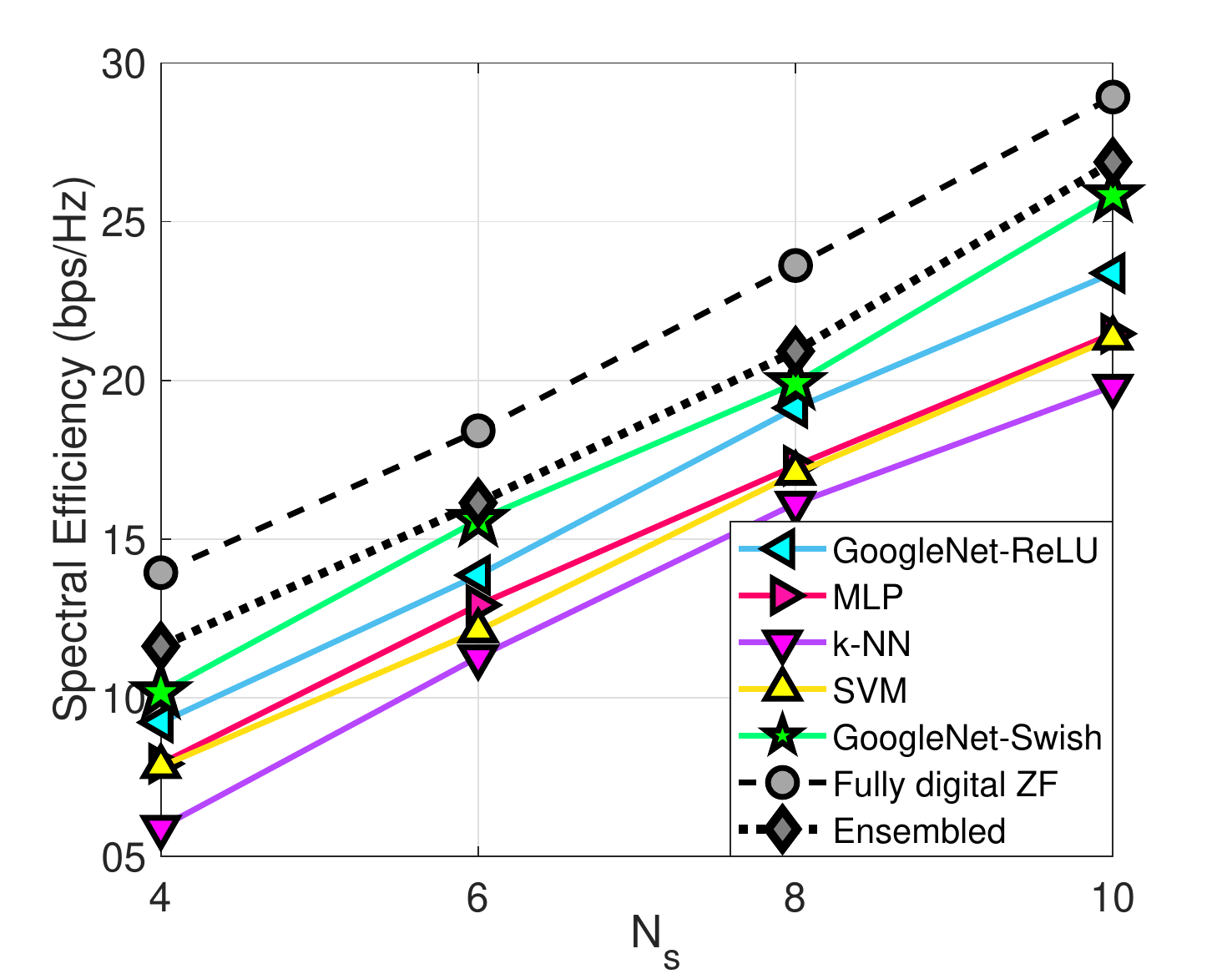}&\hspace{0.15cm}\includegraphics[width=10.80cm,height=3.90cm]{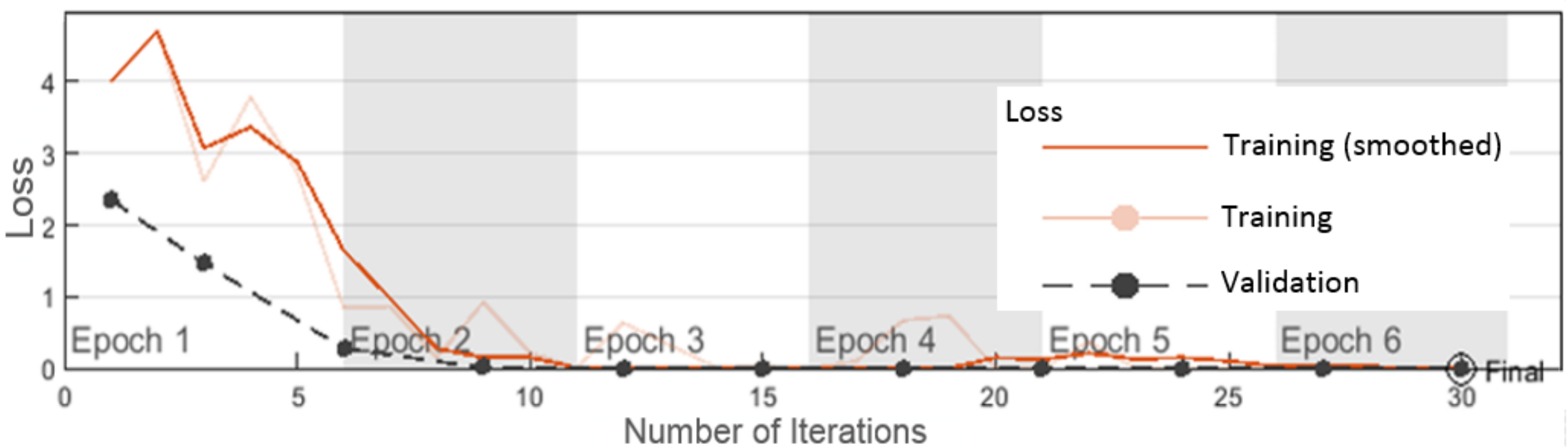}\\
	\hspace{0.40cm}(c) Achievable SE vs. varying $N_{\textrm{s}}$ & \hspace{0.3cm} (d) The convergence of the Swish-driven GoogleNet (loss)
	\end{tabular}
		\caption{The convergence and performance of the fine-tuned GoogleNet for analog beam selection.}
		\label{fig:GoogleNet}
\end{figure*}

\subsection{Enhancing Accuracy via Ensemble Learning}
We further improve the accuracy of the proposed procedure for analog beam selection through the ensemble learning technique, which puts forward to train a strong ensembled model, that combines the predictions of distinct weak learners (e.g., the Swish-driven GoogleNet modules in this paper) to achieve a more precise model. To do so, a gradient boosting (gradBoost) mechanism\cite{gradBoost} is adopted, wherein we sequentially train the weak learners. 
\par Towards forming an ensembled model as in Fig.~\ref{fig:ensemble}, we adopt $M_{1}$ random subsets $Z_{m} (m\in{M_{1}})$ of the whole training sample set $\mathcal{Z}$, where the weak learners are trained upon different subsets. For any sample $z_{m}\in{Z_{m}}$ of size $\mathbb{C}^{1\times (4N_{{cl}}\times N_{{ray}}+2)}$, the weak learner performs a classification and assigns a specific class from $\omega_m\in{\Omega=\{0,...,N^{RF}_{\textrm{t/r}}\}}$. The goal in each step is boosting the training accuracy of the current weak learner through focusing on the misclassified observations made by the previous ones. The misclassified samples are injected forward to train the next weak learner more efficiently. The strong ensembled learner thereafter adopts a majority voting mechanism based on a weighted summation of $M_1$ weak learners. To this goal, a voting counter $\Psi(\omega)\in{\mathbb{N}^{1\times\Omega}}$ indicates the number of classifiers, which adopted the RF chain class $\omega$. The weighted summation is given by $\Phi_{M_{1}}^{ens}=\sum_{m=1}^{M_1}c_{m}\Psi_{m}(\omega)$, where $c_{m}$ denotes the weight of the $m$th Swish-driven GoogleNet, indicating the performances of this weak model. Indeed, the better a weak learner performs, the more it contributes to the strong ensembled model. The strong ensembled learner thus, is generally less biased than the weak learners, since the misclassified observations are efficiently propagated and learned along the ensembling chain. 
The challenge here, is to select the optimal order of the classifiers to be trained within the ensembling chian, i.e., obtaining the optimal order of $\Phi_{M_{1}}^{ens}$ is complicated, especially for a long ensembling chain. 
\par Instead of optimizing this order globally, we are seeking for the best possible pairs of ($c_{m},\Psi_{m}(\omega)$) to be locally built and iteratively added in a sub-optimal approach. The strong ensembled model can be recurrently indicated by $\Phi_{m}^{ens}=\Phi_{m-1}^{ens}-c_{m}\nabla_{\Phi_{m-1}^{ens}} E(\Phi_{m-1}^{ens})$, whereby the best possible pair ($c_{m},\Psi_{m}(\omega)$) can be obtained as $(c_{m},\Psi_{m}(\omega)) =\argmin_{c,\Psi(\omega)}E(\Phi_{m-1}^{ens}+c\Psi(\omega))$, with $E(.)$ denoting the strong ensembled learner fitting error. Finally, the RF chain class $\omega$, which maximizes the voting counter $\Psi_{m}(\omega)$ by contributing $M_{1}$ weak learners and their impacts, is adopted by the strong ensembled learner as $\omega^{*}=\argmax_{\omega\in{\Omega}}\frac{1}{M_{1}}\sum_{m=1}^{M_{1}}c_{m}\Psi_{m}(\omega)$.
\section{Simulation Results}
We consider a clustered THz channel with $N_{{cl}}=$~4 clusters and $N_{{ray}}=$~2 propagation rays in each cluster. The signal wavelength is $\lambda = $~1.36, the AoAs and the AoDs are uniformly distributed within $[-\frac{1}{2},\frac{1}{2}]$, while the complex-valued gain follows $\mathcal{CN}$(0, 1). Simulations are performed for a lens-aided MIMO system equipped with $N_{{\textrm{r}}}=$~64, $N_{{\textrm{t}}}=$~256 and $N_{{\textrm{r}}}^{\textrm{RF}}=N_{{\textrm{t}}}^{\textrm{RF}}=$~4. For the simulations related to the GoogleNet as indicated in Table I, we used 70\% of the sampling data for the training and the rest are for the validation. Moreover, the “MiniBatchSize” shows the number of images used at each iteration of training/validation. The maximum number of training epochs is indicated by “MaxEpochs” and the “Shuffle” field is every epoch, which randomly initiates a new datastore with the same training/validation data. The
initial learning rate “InitialLearnRate” slows down the learning process, in the transferred layers owing to its adopted small value and the “ValidationFrequency” field specifies that the validation is performed every three iterations during training.
The achievable SE of a hybrid analog-digital beamspace system can be expressed as
$\!SE \!=  \textrm{log}_{2}\big|\textbf{I}_{N_{\textrm{s}}}\!+\!\frac{\rho}{\sigma^{2}N_{\textrm{s}}}R_{n}^{-1}(\textbf{W}_{\textrm{BB}})^{H}(\textbf{S}_{\textrm{r}})^{H}\textbf{H}_{{b}}\textbf{S}_{\textrm{t}}\textbf{F}_{\textrm{BB}} (\textbf{F}_{\textrm{BB}})^{H}(\textbf{S}_{\textrm{t}})^{H}(\textbf{H}_{{b}})^{H}\textbf{S}_{\textrm{r}}\textbf{W}_{\textrm{BB}}\big|,$
where $R_{n}=\!(\textbf{W}_{\textrm{BB}})^{H}(\textbf{S}_{\textrm{r}})^{H}\textbf{S}_{\textrm{r}}\textbf{W}_{\textrm{BB}}$ is the noise covariance matrix after combining.
\begin{table}
\centering
\captionsetup{font=small}
\captionsetup{justification=centering}
\caption{\\GoogleNet configurations}
\label{table:notations}
\begin{tabular}
{ | m{3.0cm} | m{2cm} | }
\hline
\textbf{Parameter} & \textbf{Value} \\ 
\hline\hline
TrainingSize & 70\% \\
\hline
ValidationSize & 30\% \\
\hline
MiniBatchSize & 128 \\
\hline
MaxEpochs & 6\\
\hline
Shuffle & every epoch \\
\hline
InitialLearnRate & 1e-3 \\
\hline
ValidationFrequency & 3 \\
\hline
\end{tabular}
\end{table}
\par The analog beam selection baseline strategies MLP, $k$-NN, and SVM with the same internal configurations in \cite{access2}, the conventional ReLU-driven GoogleNet, the modified Swish-driven GoogleNet, and the ensemble learning schemes are investigated for comparison in terms of achievable SE. Additionally, the fully digital zero-forcing (ZF) strategy by using the whole beams at the transceivers, is the optimal benchmark baseline. 
\par First, we assess the convergence accuracy and loss ratios for the training/validation process of the proposed Swish-driven GoogleNet scheme in Figs.~\ref{fig:GoogleNet}(b) and \ref{fig:GoogleNet}(d), respectively. Clearly, the training/validation process is inaccurate in first iterations. That is because the weights and biases of the input layer and the linear layer are not well fine-tuned with the sampling data. Gradually as the iterations progress, the training/validation accuracy improves (tends to 100\%), while the training/validation loss degrades (tends to 0). 
\par Next, we analyze the performance of our proposed schemes in a comparative fashion. The benchmark fully-digital ZF strategy with $N^{RF}_{\textrm{t}}=$~256 and $N^{RF}_{\textrm{r}}=$~16 RF chains obviously, has the largest achievable SE in Fig.~\ref{fig:GoogleNet}(a) and  Fig.~\ref{fig:GoogleNet}(c) at the expense of severe system complexity, energy consumption, and hardware cost. Fig.~\ref{fig:GoogleNet}(a) with varying SNR in 0dB$\sim$30dB and  $N_{\textrm{t}}^{RF}=N_{\textrm{r}}^{RF}=N_{\textrm{s}}$, where $N_{\textrm{s}}=$~4, indicates that by increasing the SNR, the achievable SE improves for all the baselines. According to  Fig.~\ref{fig:GoogleNet}(c) with varying $N_{\textrm{s}}$ in 4$\sim$10, where $N_{\textrm{t}}^{RF}=N_{\textrm{r}}^{RF}=N_{\textrm{s}}$ and SNR~=~10dB, the achievable SE increases for more number of simultaneous data streams. Our proposed ensemble learning scheme is the most superior amongst others and is the closest scheme to the benchmark due to a better accuracy. This scheme according to Fig.~\ref{fig:GoogleNet}(a), improves the achievable SE of the MLP scheme \cite{access2} at SNR~=~30dB, by up to 27\%. Similarly at SNR~=~30dB, the proposed Swish-enabled GoogleNet and the conventional ReLU-driven GoogleNet schemes achieve a better performance than other strategies MLP, SVM, and $k$-NN, by exhibiting 18\% and 12\% achievable SE gain compared to the MLP scheme\cite{access2}, respectively.
\par
\begin{figure}
\centering
\hspace{0.0cm}\includegraphics[width=6.0cm,height=4.50cm]{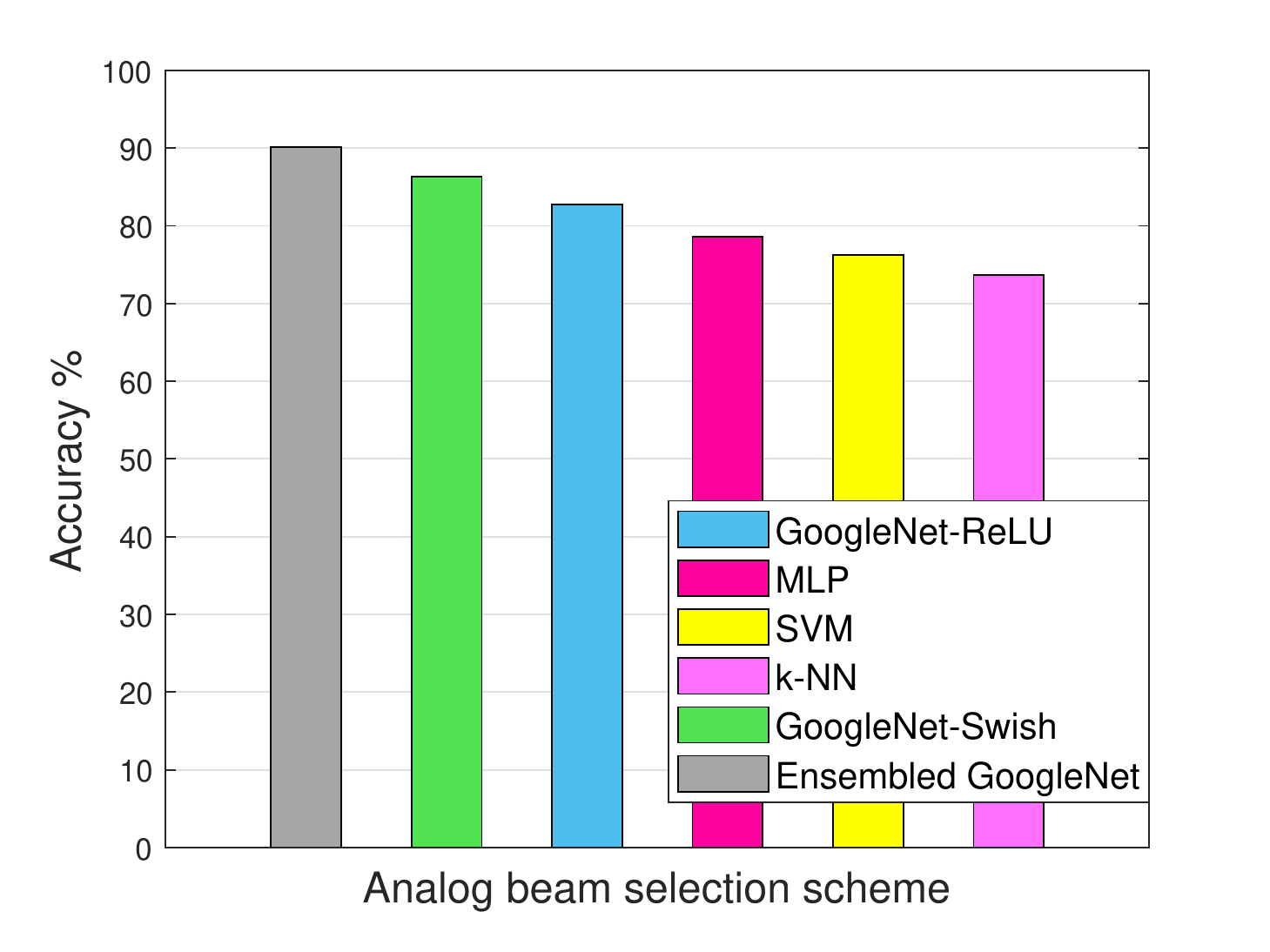}\caption{Aanalog beam selection accuracy comparison.}
\label{fig:acc}
\end{figure}

 In Fig.~\ref{fig:acc} under the same configurations in Fig.~\ref{fig:GoogleNet}(c) with $N_{\textrm{s}}=$~4, the accuracy of the analog beam selection strategies is assessed. The ensemble learning strategy with 90\% accuracy is the best, while the Swish-driven GoogleNet and the conventional ReLU-driven GoogleNet schemes with 86\% and 83\% on average, are the second and third best strategies for analog beam selection. The reason is that retraining/modifying the pre-trained networks such as GoogleNet based on transfer learning for the classification tasks (e.g., analog beam selection) is more accurate than training a deep network such as MLP\cite{access2} from scratch. 
Inspired by the transfer learning method, the parameters in a pre-trained deep structure are mostly kept unchanged, while few certain parameters are fine-tuned based on samples.
We further examine the accuracy of the conventional ReLU-driven GoogleNet, as well as the fine-tuned Swish-driven GoogleNet schemes by applying different training functions e.g., root mean square propagation (RMSPROP), adaptive moment estimation (ADAM) and stochastic gradient descent method (SGDM), as demonstrated in Table II. One can observe that the Swish-driven GoogleNet scheme trained by the SGDM can achieve the best analog beam selection accuracy.
\begin{table}
\captionsetup{font=small}
\captionsetup{justification=centering}
	\caption{\\GoogleNet-based analog beam selection accuracy comparison.}
	\begin{center}
		\begin{tabular}{ |m{2.5cm}| m{1.2cm}| m{1cm} |m{1cm}| }
			\hline
			\textbf{Architecture/Function} & {RMSPROP} & {ADAM} & {SGDM} \\ 
			\hline\hline
			GoogleNet-ReLU & 83.4\% & 81.37\% & 82.22\% \\
			\hline
			GoogleNet-Swish & 86.21\% & 85.27\% & 86.93\% \\
			\hline
		\end{tabular}
		\label{table:comp}
	\end{center}
\end{table}
\section{Conclusions}
In this paper, we proposed a novel deep learning technique framework to address the analog beam selection problem in a THz beamspace MIMO system. Specifically, we retrained the pre-trained off-the-shelf GoogleNet for learning the analog beam selection based on the concept of transfer learning. Then, we fine-tuned the GoogleNet enabling the Swish activation function, for a better analog beam selection precision. Finally, an ensemble learning technique presented for boosting the precision beyond a conventional fine-tuned GoogleNet. Simulations revealed a remarkable enhancement in accuracy, as well as in achievable SE.
\section*{Acknowledgement}
This work is supported by the Academy of Finland: (a) ee-IoT n.319009, (b) EnergyNet n.321265/n.328869, and (c) FIREMAN n.326270/CHISTERA-17-BDSI-003; and by JAES Foundation via STREAM project.

\end{document}